# ON THE POSSIBILITY OF THz AMPLIFICATION AND GENERATION IN NANO-STRUCTURES, BY MEANS OF 2D PLASMA WAVES

A.V. Chaplik [1,2]   and   G.P. Berman [1]

[1]) Theoretical Division, MS B213, Los Alamos National Laboratory, Los Alamos, New Mexico 87545

[2]) Institute of Semiconductor Physics, Siberian Division of RAS, Novosibirsk, 630090, Russia

Plasma oscillations in 2D systems are characterized by a rather complex and technologically controllable dispersion law (DL). Freely hanging electron sheet supports the plasma waves with the square root DL: $\omega \sim k^{1/2}$, where $\omega$ is the frequency and $k$ is the wave vector [1]. In gated structures, like MOS (metal – oxide – semiconductor), the square root dispersion appears when $k$ increases after a linear DL domain, and the effective "sound velocity" of the plasmon depends on the thickness of oxide layer [2]. Besides, there are two other linear DL domains: at very small $k$ due to the retardation effects (the slope of $\omega(k)$ tends to the speed of light) and at very large $k$ - zero sound regime (the slope equals the Fermi-velocity).

Experimentally typical are the plasmon wavelengths of about one micron [3]. For the density of 2D electrons $10^{12}$ /cm$^2$ (also experimentally typical) the corresponding frequency of 2D plasma oscillations lies in THz band.

The idea to use the 2D plasmons for generation of THz radiation was first formulated in [4] (see also [5]), where a MOS-structure on silicon was considered. The authors have shown that in the linear DL regime a rather short device (a few tenths of microns) possesses a specific type of instability that is similar to the so called "whistle" instability of acoustic systems. A definite value of $k$ is fixed just by the length of the device. Due to its small size (much less than electro-magnetic (EM) wavelength of the corresponding frequency) such a device operates as a practically point source of THz radiation.

Recent interest in 2D plasmons has been stimulated by the technological successes in fabrication of the so called double-quantum-well structures. These are two parallel thin films (~100 – 200 A) of a narrow band semiconductor (e.g. GaAs, InAs) separated by a layer of material with a wider forbidden gap (GaAlAs ). In such a structure one deals with two components of 2D plasma coupled both electrically (by means of the electrostatic forces) and electronically (due to charge carriers tunneling through the barrier formed by the wide-gap material). The plasma -wave spectrum of these systems consists of a few branches [6]. The co-phase oscillations are known as optical plasmons while the anti-phase oscillations form the acoustic plasmon branch (linear DL at small $k$- vectors). Both of these branches correspond to density fluctuations of electrons (each group in its "own" quantum well) without interlayer transitions of the particles (purely electrostatic coupling). The branch associated with the interchange of electrons between different components of 2D plasma is called the " intersubband plasmon" because with tunneling allowed there is a potential well having two minima. Spreading the electrons over the entire well results in a two-level system (a two-subband system if one takes into account the free motion of electrons in X-Y plane while Z-axis determines the growth direction of the structure).

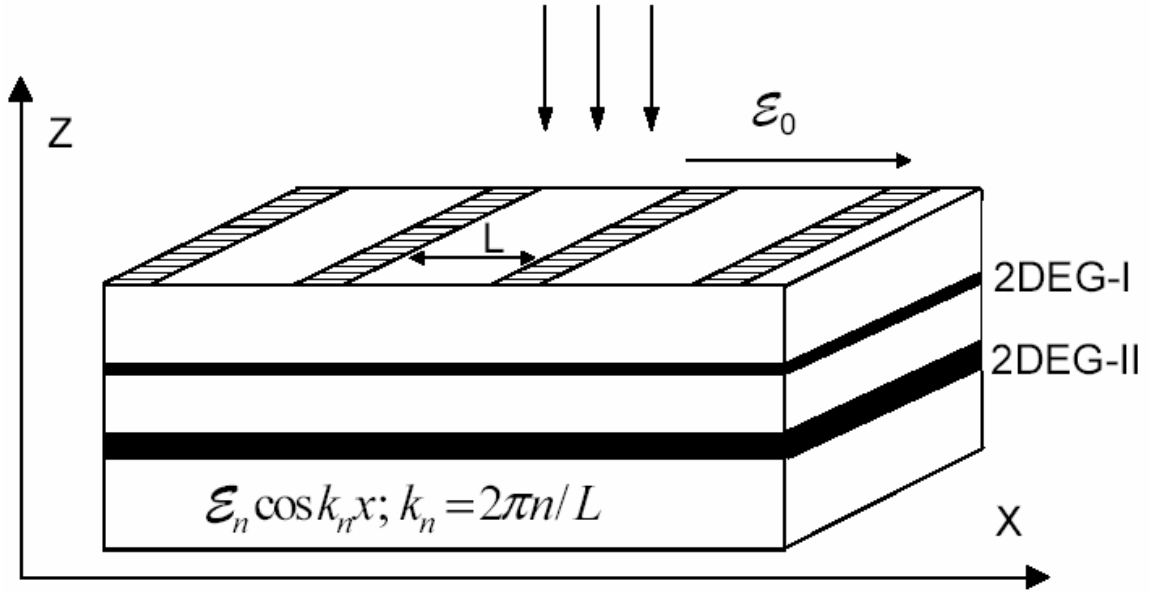

Fig. 1: Schematic of the proposed structure. Metal stripes on the top form the grating structure with the period L; $\mathcal{E}_0$ - electric field of the incoming EM wave (vertical arrows); 2DEG-I and 2DEG-II - 2D electrons in the adjacent quantum wells; $k_n$ and $\mathcal{E}_n$ - wave number and the amplitude of the n-th Fourier component of the electric field, respectively.

If the system in question is symmetric with respect to its middle plane, the wave functions of the transversal motion are characterized by parity. In this case all three branches are decoupled and, for appropriate values of the areal density of electrons and energy separation between single-particle levels, the optical and intersubband branches can cross. Asymmetry of the structure results in the mixing and anticrossing of those modes as has been shown in [7,8]. In an asymmetric two-component system a specific mechanism of the plasmon damping occurs and the amplification of plasma waves becomes possible at the inversion of the population of the transverse subbands (see [9]). This idea forms the basis of the present proposal.

Thus, consider the bilayer structure (double-quantum-well) depicted in Fig. 1. Two bold lines represent 2D electron gases (2DEG) located in adjacent quantum wells. The potential energy of an electron in the structure is shown in Fig. 2 along with two lowest levels of the transversal quantization. The grating structure on the top of the specimen transforms the electric field of the incident EM wave polarized perpendicularly to the metal strips in such a way that the electrons in 2DEG-I and 2DEG-II "see" a spatially modulated electric field with periods L/n, n = 1,2,3…..This allows one to "introduce" a definite wave vector for the 2D plasmons in the system [10]. Also, the same grating structure acts as an emitting antenna that converts the fields accompanying plasma oscillations into EM radiation with much better angular characteristics than a point source.

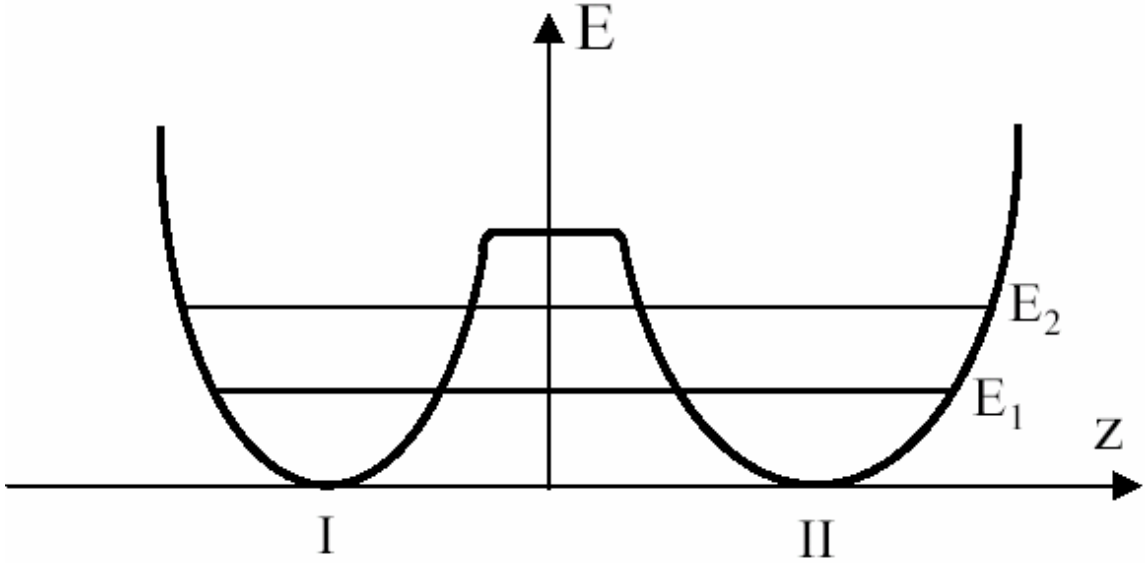

Fig. 2: Electron potential energy in an asymmetric double-quantum well. $E_1$ and $E_2$ - energy levels of the transverse motion.

The plasma waves decay (apart from the trivial collision mechanisms) due to Landau damping, which is nothing but the reciprocal Cherenkov effect: each plasmon fathers an electron-hole pair or, in other words, the energy – momentum conservation laws allow a single particle transition; the energy transfer and the momentum transfer in such a process equal the energy and momentum of the plasmon. For our system the corresponding equation reads:

$$\hbar\omega(\mathbf{k}) = E_2(\mathbf{p}+\mathbf{k}) - E_1(\mathbf{p}),$$

where $E_{1,2}(\mathbf{p})$ are the single particle energies in the subbands 1 and 2 with an electron momentum $\mathbf{p}$, $\omega$ and $\mathbf{k}$ are the frequency and the wave vector of the plasmon. Thus, we deal with a specific version of Landau damping in which one of the components of the electron momentum is quantized. The energy of the wave is transferred to the excitation of the "oblique"($\mathbf{p}$ is not conserved) intersubband transitions in accordance with the equation above. However, the process allowed by the energy –momentum conservation law is not necessarily realized due to symmetry reasons. For example, the optical plasmons in symmetric structures are accompanied by an electric field that is even when reflecting in the middle plane. That is why the intersubband transitions are forbidden by the parity conservation law. However, in asymmetric structures (different widths of the wells, different electron concentrations) such transitions become possible and we get a new mechanism of the collisionless absorption for incoming EM waves with THz frequencies.

Calculations show that (as it must be !) the absorbed power, Q, is proportional to the factor of asymmetry ( defined by the electron wave functions of the transversal motion) and proportional to the quantity $(N_1 - N_2)$ - the difference of the subband populations. Hence, if an inverted population of subbands is created by external pumping, the quantity Q becomes negative and one gets amplification of the optical plasma wave (provided the modulus Q exceeds the losses, of course). The inverted population of the subbands can be achieved in the process of the vertical (tunneling) transport through the bilayer structure (pumping by current) or by the optical excitation of electrons from the valence band (optical pumping).

Numerical estimates have been made for room temperature and for sufficiently good (but not of record quality) GaAs specimen: the mobility at liquid He temperatures is $10^6$ cm$^2$/Vs; at room temperature it is 8000 cm$^2$/Vs. The mobility gives us an estimate for the collision induced damping (losses). Then, assuming the real density of electrons is $N_1 \sim N_2 = 10^{12}$/cm$^2$ and the inversion of the population $N_2 - N_1 = 10^{10}$/cm$^2$ (1% of the average population), we obtained the coefficient of amplification for ONE passage through the structure at the level of 4% (the ratio of the transmitted power $Q_t$ to the incident one $Q_i$ is $Q_t/Q_i = 1.04$). In this estimate, we assumed the efficiency of the grating structure is 10% for the first Fourier harmonics. This means that the amplitude of the electric field with n=1 (see Fig. 1) equals roughly 1/3 of that of incident electric field, $\mathcal{E}_0$.

Because any amplifier + positive feed back = generator, we are actually proposing a laser operating at THz frequencies. To do this, one has to place the structure in question between two mirrors parallel to the layers and possessing sufficiently good reflection coefficient in the THz band.

This work was supported by the Department of Energy (DOE) under Contract No. W-405-ENG-36.